\documentclass[reprint,
 amsmath,amssymb,
 aps,
pra,
]{revtex4-1}
\usepackage{dcolumn}
\usepackage{bm}
\usepackage{amsfonts}
\usepackage{amsmath}
\usepackage{amssymb}
\usepackage{float}
\usepackage{graphicx}
\usepackage{subfigure}
\usepackage{epstopdf}
\usepackage{hyperref}
\usepackage{enumerate}
\usepackage{verbatim}
\newcommand{\ket}[1]{|#1\rangle}

\begin{document}
\title{Eta-pairing states in Hubbard models  with bond-charge interactions on general graphs}
\author{Ming-Yong Ye$^{1,2,}$}
\email{myye@fjnu.edu.cn}
\affiliation{$^{1}$College of Physics and Energy, Fujian Normal University, Fujian Provincial Key Laboratory of Quantum Manipulation and New Energy Materials, Fuzhou, 350117, China.}
\affiliation{$^{2}$Fujian Provincial Engineering Technology Research Center of Solar Energy Conversion and Energy Storage, Fuzhou, 350117, China.}

\begin{abstract}
We investigate Hubbard models with bond-charge interactions on general graphs. For a Hamiltonian \(H\) of such a model, we provide the condition on its parameters under which the \(\eta\)-pairing method can be employed to construct its exact eigenstates. We arrive at this condition by finding that the requirement for the \(\eta\)-pairing state \((\eta^\dagger)^N |0\rangle\) to be an eigenstate of \(H\) is identical to the requirement for it to be an eigenstate of a Hubbard-type Hamiltonian \(H_m\). When the condition for \((\eta^\dagger)^N |0\rangle\) to be an eigenstate of the Hubbard-type Hamiltonian \(H_m\) is satisfied, we demonstrate that there are additional states, distinct from \((\eta^\dagger)^N |0\rangle\), which are also exact eigenstates of \(H_m\).
Our results enhance the understanding of Hubbard models on general graphs, both with and without bond-charge interactions. 
\end{abstract}

\maketitle

\section{introduction}
The Hubbard model provides a framework for investigating strongly correlated electron systems \cite{hubbard}, particularly in two dimensions where it gives valuable insights into the mechanisms underlying high-temperature superconductivity \cite{annurev:/content/journals/10.1146/annurev-conmatphys-031620-102024}.
Its Hamiltonian  has two terms: the first term describes the hopping of electrons between different sites, and the second term describes the interaction when two electrons are located at the same site. Yang makes a pioneering work to the Hubbard model on a hypercubic lattice by
 constructing some exact eigenstates of the model \cite{PhysRevLett.63.2144}.
 His approach relies on the introduction of the $\eta$-pairing operator, which serves as an eigenoperator of the system's Hamiltonian.
There are also some efforts to find new exact eigenstates of the Hubbard model using other methods \cite{myye}.
The $\eta$-pairing method has been successfully extended to construct exact eigenstates for certain Hubbard models on triangular lattices \cite{PhysRevB.71.012512}. More recently, the conditions under which the $\eta$-pairing method can effectively identify exact eigenstates for Hubbard models on general graphs have been explored \cite{PhysRevB.102.085140}, and
there are discussions about the $\eta$-pairing method
on non-Hermitian Hubbard models \cite{PhysRevB.103.235153,arxivlikai} and multicomponent Hubbard models \cite{PhysRevResearch.6.043259}.

To further enrich the Hubbard model and explore related physics, additional interactions can be incorporated. One such interaction is the bond-charge interaction, which modifies the hopping strengths depending on the occupation across the bond \cite{PhysRevB.39.11515}. The inclusion of bond-charge interactions in the Hubbard model has been used as a model of hole superconductivity \cite{PhysRevB.39.11515}, and some exact eigenstates for this extended model have also been constructed \cite{myye}.
Quantum many-body scar states have received many attentions in recent years, as these states do not obey the eigenstate thermalization hypothesis \cite{PhysRevA.43.2046,PhysRevE.50.888}.
It has been demonstrated that the Hubbard models with bond-charge interactions on hypercubic lattices can support $\eta$-pairing states as their eigenstates \cite{PhysRevB.102.075132}. These states are true scar states, as the bond-charge interaction breaks the $\eta$-pairing SU(2) symmetry inherent in the original Hubbard models \cite{PhysRevB.102.075132}.

In this paper,
we give the condition for the $\eta$-pairing state $(\eta^\dag)^{N}\ket{0}$ to be an eigenstate of a Hubbard model with bond-charge interaction on a general graph, which is based on a finding that
 the condition for the $\eta$-pairing state $(\eta^\dag)^{N}\ket{0}$ to be an eigenstate of a Hubbard model with bond-charge interaction is identical to the condition for it to be an eigenstate of a modified Hubbard model without bond-charge interactions.
We also identify several eigenstates of the Hubbard model that are distinct from the $\eta$-pairing state $(\eta^\dag)^{N}\ket{0}$.

\section{Condition for  $\eta$-pairing states to be eigenstates}
Consider a Hubbard model  with bond-charge interaction defined on a general graph,
the Hamiltonian is
\begin{align} \label{H}
H=&\sum_{\langle m,n\rangle} \left \{T_{mn} +B_{mn}  \right \} +U\sum_n a_n^\dag a_n b_n^\dag b_n,
\end{align}
where $m$ and $n$ denote sites on the graph, $a_n$ and $b_n$ are annihilation operators for spin-up and spin-down electrons on site $n$, and $a_n^\dag$ and $b_n^\dag$
are the corresponding creation operators. The ordinary  hopping $T_{mn}$ is
\begin{align}
&T_{mn}=t_{mn}(a_m^\dag a_n+b_m^\dag b_n)+t_{mn}^*(a_n^\dag a_m+b_n^\dag b_m).
\end{align}
The bond-charge interaction  $B_{mn}$ is
\begin{align}
B_{mn} =&( \chi_{mn}a_m^\dag a_n+\chi_{mn}^*a_n^\dag a_m)(b_m^\dag b_m+b_n^\dag b_n) \nonumber \\
&+(\chi_{mn} b_m^\dag b_n+\chi_{mn}^*b_n^\dag b_m)(a_m^\dag a_m+a_n^\dag a_n).
\end{align}
The notation $\langle m,n \rangle$ means the sum is over the pair of sites connected on the graph.

Divide the bond-charge interaction $B_{mn}$ as $B_{mn}=B_{mn1}+B_{mn2}$ with
\begin{equation}
B_{mn1}=\chi_{mn}(a_m^\dag a_n+b_m^\dag b_n)+\chi_{mn}^*(a_n^\dag a_m+b_n^\dag b_m),
\end{equation}
and
\begin{align}
B_{mn2}&= \chi_{mn}a_m^\dag a_n(b_m^\dag b_m-b_n b_n^\dag )   \nonumber \\
       &+\chi_{mn}^*a_n^\dag a_m(b_n^\dag b_n-b_m b_m^\dag) \nonumber \\
       &+\chi_{mn} b_m^\dag b_n (a_m^\dag a_m-a_n a_n^\dag )\nonumber \\
       &+\chi_{mn}^*b_n^\dag b_m(a_n^\dag a_n-a_m a_m^\dag). \label{xa1}
\end{align}
We can write
\begin{equation}
T_{mn}+B_{mn}=(T_{mn}+B_{mn1})+B_{mn2},
\end{equation}
where $T_{mn}+B_{mn1}$ has the same form as the hopping $T_{mn}$ but with the strength $t_{mn}+\chi_{mn}$.
So adding the bond charge interaction $B_{mn}$ to the Hubbard model can be understood as changing the hopping strength from $t_{mn}$ to $t_{mn}+\chi_{mn}$, and
adding the leftover interaction $B_{mn2}$.

Now we discuss the condition for the  $\eta$-pairing states to be eigenstates of the Hamiltonian $H$.
The $\eta$-pairing operator is defined as
\begin{equation}
\eta^\dag=\sum_n \eta_n^\dag,\quad  \eta_n^\dag=e^{i\phi_n}a_n^\dag   b_n^\dag. \label{etax}
\end{equation}
There is a good relation between the operator $B_{mn2}$ and $\eta^\dag$, i.e.,
\begin{align}
&[[B_{mn2},\eta^\dag],\eta^\dag]= 0, \label{x100}
\end{align}
which is proved as follows.
In the following $n \ne m$, simple calculations show that there are
\begin{align}
&[a_m^\dag a_nb_m^\dag b_m,e^{i\phi_{m}}a_m^\dag b_m^\dag]=0,\\
&[a_m^\dag a_nb_m^\dag b_m,e^{i\phi_{n}}a_n^\dag b_n^\dag]=e^{i\phi_{n}}a_m^\dag b_n^\dag b_m^\dag b_m,\\
&[-a_m^\dag a_nb_n b_n^\dag,e^{i\phi_{m}}a_m^\dag b_m^\dag]=0,\\
&[-a_m^\dag a_nb_n b_n^\dag,e^{i\phi_{n}}a_n^\dag b_n^\dag]=-e^{i\phi_{n}}a_m^\dag b_n^\dag a_n^\dag a_n.
\end{align}
The sum of the above four equations leads to
\begin{align}
&[a_m^\dag a_n(b_m^\dag b_m-b_n b_n^\dag),\eta_m^\dag+\eta_n^\dag] \nonumber  \\
&=e^{i\phi_{n}}a_m^\dag b_n^\dag (b_m^\dag b_m-a_n^\dag a_n). \label{x4}
\end{align}
In Eq. (\ref{x4}), interchanging $m$ and $n$ leads to
\begin{align}
&[a_n^\dag a_m(b_n^\dag b_n-b_m b_m^\dag),\eta_m^\dag+\eta_n^\dag] \nonumber  \\
&=e^{i\phi_{m}}a_n^\dag b_m^\dag (b_n^\dag b_n-a_m^\dag a_m).
\end{align}
In Eq. (\ref{x4}), interchanging $a$ and $b$ leads to
\begin{align}
&[b_m^\dag b_n(a_m^\dag a_m-a_n a_n^\dag),\eta_m^\dag+\eta_n^\dag] \nonumber  \\
&=e^{i\phi_{n}}a_n^\dag b_m^\dag  (a_m^\dag a_m-b_n^\dag b_n). \label{x5}
\end{align}
In Eq. (\ref{x5}), interchanging $m$ and $n$ leads to
\begin{align}
&[b_n^\dag b_m(a_n^\dag a_n-a_m a_m^\dag),\eta_m^\dag+\eta_n^\dag] \nonumber  \\
&=e^{i\phi_{m}}a_m^\dag b_n^\dag  (a_n^\dag a_n-b_m^\dag b_m).  \label{eq42}
\end{align}
Notice the expression of $B_{mn2}$ in Eq. (\ref{xa1}), a linear combination of the above four equations leads to
\begin{align}
&[B_{mn2},\eta^\dag]=[B_{mn2},\eta_m^\dag+\eta_n^\dag]=g_{mn} M_{mn}, \label{x6}
\end{align}
where
\begin{align}
g_{mn}=&\chi_{mn} e^{i\phi_{n}}- \chi_{mn}^*e^{i\phi_{m}},\\
M_{mn}=&a_m^\dag b_n^\dag(b_m^\dag b_m-a_n^\dag a_n) \nonumber \\
       &+a_n^\dag b_m^\dag(a_m^\dag a_m-b_n^\dag b_n).
\end{align}
As there is
\begin{equation}
 [M_{mn},\eta^\dag]=[M_{mn},\eta^\dag_m]+[M_{mn},\eta^\dag_n]=0,
\end{equation}
the relation in Eq. (\ref{x100}) is proved.

The relation in Eq. (\ref{x100}) is important as it can lead to
\begin{equation}
[B_{mn2},\eta^\dag](\eta^\dag)^N\ket{0}=(\eta^\dag)^N[B_{mn2},\eta^\dag]\ket{0}=0,
\end{equation}
where $N$ is an integer and $\ket{0}$ is the vacuum state. The above equation means that there is
\begin{equation}
B_{mn2}(\eta^\dag)^{N+1}\ket{0}=\eta^\dag B_{mn2}(\eta^\dag)^{N}\ket{0},
\end{equation}
which itself can be used $N$ times to get
\begin{equation}
B_{mn2}(\eta^\dag)^{N+1}\ket{0}=(\eta^\dag)^{N+1}B_{mn2}\ket{0}=0.
\end{equation}
Therefore the $\eta$-pairing state $(\eta^\dag)^{N}\ket{0}$ is an eigenstate of $B_{mn2}$ with zero eigenvalue.
Define
\begin{align}
H_m=&\sum_{\langle m,n\rangle} \left \{T_{mn} +B_{mn1}  \right \} +U\sum_n a_n^\dag a_n b_n^\dag b_n, \label{hm}
\end{align}
which is just the Hamiltonian of a Hubbard model with the hopping strength $t_{mn}+\chi_{mn}$ and without bond-charge interactions. As there is
\begin{equation}
H=H_m+\sum_{\langle m,n\rangle} B_{mn2},
\end{equation}
if the $\eta$-pairing state $(\eta^\dag)^{N}\ket{0}$ is an eigenstate of $H$ it will also be an eigenstate of $H_m$ and vice versa.
We can draw the conclusion that the condition for the $\eta$-pairing state $(\eta^\dag)^{N}\ket{0}$ to be an eigenstate of $H$ is the same as
that  for it to be an eigenstate of $H_m$.

To find the condition for the $\eta$-pairing state $(\eta^\dag)^{N}\ket{0}$ to be an eigenstate of $H_m$, it
 is needed to calculate the communicator $[H_m,\eta^\dag]$ as Yang's method \cite{PhysRevLett.63.2144}. The on-site interactions of $H_m$ satisfy
\begin{align}
&[U\sum_n a_n^\dag a_n b_n^\dag b_n, \eta^{\dagger}]=U\eta^{\dagger}, \label{u}
\end{align}
which can be easily verified.
To calculate the communicator $[T_{mn}+B_{mn1},\eta^\dag]$, we note that there is
\begin{equation}
[T_{mn}+B_{mn1},\eta^\dag]=[T_{mn}+B_{mn1},\eta_m^\dag+\eta_n^\dag].
\end{equation}
Direct calculation shows that there is
\begin{equation}
[a_m^\dag a_n+b_m^\dag b_n,\eta_m^\dag+\eta_n^\dag]=e^{i\phi_{n}}(a_m^\dag b_n^\dag+a_n^\dag b_m^\dag). \label{x1}
\end{equation}
In Eq. (\ref{x1}), interchanging $m$ and $n$ leads to
\begin{equation}
[a_n^\dag a_m+b_n^\dag b_m,\eta_m^\dag+\eta_n^\dag]=e^{i\phi_{m}}(a_n^\dag b_m^\dag+a_m^\dag b_n^\dag). \label{x2}
\end{equation}
A linear combination of Eq. (\ref{x1}) and Eq. (\ref{x2})  gives
\begin{equation}
[T_{mn}+B_{mn1},\eta^\dag]=f_{mn}(a_n^\dag b_m^\dag+a_m^\dag b_n^\dag),\label{x3}
\end{equation}
with
\begin{equation}
f_{mn}=(t_{mn}+\chi_{mn}) e^{i\phi_{n}}+(t_{mn}+\chi_{mn})^* e^{i\phi_{m}}.
\end{equation}
Based on the results in Eqs. (\ref{u}, \ref{x3}), we can make the summary that
\begin{align}
&[H_m,\eta^\dag]=U\eta^{\dagger}+\sum_{\langle m,n\rangle}f_{mn}(a_n^\dag b_m^\dag+a_m^\dag b_n^\dag).
\end{align}
When there is
\begin{equation}
f_{mn}=0, \quad \forall \,  \langle m,n\rangle, \label{cc}
\end{equation}
we have $[H_m,\eta^\dag]=U\eta^{\dagger}$, which can lead to
\begin{equation}
H(\eta^\dag)^{N}\ket{0}=H_m(\eta^\dag)^{N}\ket{0}=NU(\eta^\dag)^{N}\ket{0},
\end{equation}
so that  $(\eta^\dag)^{N}\ket{0}$ is an eigenstate of $H_m$ and $H$ with the eigenvalue $NU$.
Therefore Eq. (\ref{cc}) is
 the condition for the $\eta$-pairing state $(\eta^\dag)^{N}\ket{0}$ to be an eigenstate of $H_m$ and $H$.

The condition in Eq. (\ref{cc}) gives a constraint on the hopping strength $t_{mn}$, the bond-charge interaction strength $\chi_{mn}$  and the phases $ e^{i\phi_{n}}$ in $\eta^{\dagger}$.
In the special case that $t_{mn}+\chi_{mn}=0$ for all pairs $\langle m,n \rangle$, Eq. (\ref{cc}) gives no requirement on the phases in $\eta^{\dagger}$,
so $\eta$-pairing states with any phases in $\eta^{\dagger}$ are eigenstates of $H$. Such a special case is studied in detail  in one dimension \cite{PhysRevLett.73.2240,PhysRevB.51.10386}.
In the following we assume that $t_{mn}+\chi_{mn} \ne 0$, and
define
\begin{equation}
t_{mn}+\chi_{mn}=\tilde{t}_{mn}e^{i\theta_{mn}}, \label{e33}
\end{equation}
where $\tilde{t}_{mn}$ and $\theta_{mn}$ are real. Now the condition in Eq. (\ref{cc})  for the $\eta$-pairing state $(\eta^\dag)^{N}\ket{0}$ to be an eigenstate of $H_m$ and $H$ is equivalent to
\begin{equation}
e^{2i\theta_{mn}}=-e^{i\phi_{m}}e^{-i\phi_{n}}, \quad \forall \,  \langle m,n\rangle. \label{c}
\end{equation}
which will be discussed in the following from two different views.

\section{When the phases in $\eta^{\dagger}$ are specified}
Based on the condition in Eq. (\ref{c}), we ask the following question:
assume that the phases in $\eta^{\dagger}$ for all sites are specified, can we find a Hubbard model with  bond-charge interaction such that the
state $(\eta^\dag)^{N}\ket{0}$ is an eigenstate of $H$? The answer is always YES. Such a model can be achieved when the phase of $t_{mn}+\chi_{mn}$ is
\begin{equation}   \label{c1}
e^{i\theta_{mn}}=e^{\frac{1}{2}i(\phi_m-\phi_n\pm\pi)}, \quad \forall \,  \langle m,n\rangle,
\end{equation}
where $\phi_m$ and $\phi_n$ are  phases in $\eta^{\dagger}$,
as it satisfies the condition in Eq. (\ref{c}).

Two interesting examples can be used to demonstrate this result.
In the first example, define
\begin{equation}
\eta_1^\dag=\sum_na_n^\dag b_n^\dag,
\end{equation}
which takes $\phi_n=0$ for all $n$. If we want $(\eta_1^\dag)^{N}\ket{0}$  to be an eigenstate of $H$, according to Eq. (\ref{c1})
we can set $e^{i\theta_{mn}}=\pm i$, which means  $t_{mn}+\chi_{mn}$ are all pure imaginary numbers.
In the second example, define
\begin{equation}
\eta_2^\dag=\sum_{n\in A} a_n^\dag b_n^\dag-\sum_{n\in B} a_n^\dag b_n^\dag, \label{eta2}
\end{equation}
where the sets of sites $A$ and $B$ form a bipartition of the graph.
In this example there is $\phi_n=0$ for all $n\in A$ and $\phi_n=\pi$ for all $n\in B$.
If we want $(\eta_2^\dag)^{N}\ket{0}$  to be an eigenstate of $H$, according to Eq. (\ref{c1})  we can set $e^{i\theta_{mn}}=\pm i$
when both $m,n$ are in $A$ or $B$, and $e^{i\theta_{mn}}=\pm 1$ when $n\in A, m\in B$, or $m\in A, n\in B$.
This requirement means that
the hopping strength $t_{mn}+\chi_{mn}$ within $A$ or $B$ are all pure imaginary numbers, and between $A$ and $B$ are real numbers.
Therefore, for a  Hubbard model defined on square lattices with nearest-neighbour hoppings and
next-nearest-neighbour hoppings, if the nearest-neighbour hopping coefficients are real numbers and the next-nearest-neighbour hopping
coefficients are pure imaginary numbers, the $\eta$-pairing state $(\eta_2^\dag)^{N}\ket{0}$  will be an eigenstate of the model Hamiltonian.

\section{When the parameters in $H$ are specified}
Based on the condition in Eq. (\ref{c}), we can also ask the following question:
assume that the parameters in $H$ are specified, can we find an eta-pairing operator $\eta^\dag$ such that the
state $(\eta^\dag)^{N}\ket{0}$ is an eigenstate of $H$? The answer depends on the  phase $\theta_{mn}$  of $t_{mn}+\chi_{mn}$.

Suppose sites $(n_1,n_2,\cdots,n_k,n_1)$ form a loop on the graph,
when the condition in Eq. (\ref{c}) is satisfied,  there is
\begin{equation}
e^{2i\theta_{n_2n_1}}e^{2i\theta_{n_3n_2}}\cdots e^{2i\theta_{n_1n_k}}=(-1)^k. \label{new}
\end{equation}
 Conversely if the condition in Eq. (\ref{new}) is satisfied for any loop $(n_1,n_2,\cdots,n_k,n_1)$, we can always find proper $e^{i\phi_n}$ in $\eta^{\dagger}$ for any site $n$ so that
the condition in Eq. (\ref{c}) can be satisfied and thus there is $[H_m,\eta^\dag]=U\eta^\dag$.
The method to find such $e^{i\phi_n}$  in $\eta^{\dagger}$ is as follows.
First set $\phi_{n_1}=0$ for an arbitrarily chosen site $n_1$. Then according to Eq. (\ref{c}) we set $e^{i\phi_m}=-e^{i\phi_{n_1}}e^{2i\theta_{mn_1}}$ for any site $m$ connected to site $n_1$.  And then  according to Eq. (\ref{c}) we set $e^{i\phi_k}=-e^{i\phi_{m}}e^{2i\theta_{km}}$ for any site $k$ connected to site $m$. Because the condition in Eq. (\ref{new}) is assumed to be satisfied for any loop,
we can get a unique $e^{i\phi_n}$ for any site $n$ by repeating this procedure.

Now we can answer the question raised at the beginning of this section. For a given $H$ with hopping strength $t_{mn}$ and bond-charge interaction strength $\chi_{mn}$, if the  phase $\theta_{mn}$  of $t_{mn}+\chi_{mn}$ satisfies the condition in Eq. (\ref{new})  for any loop $(n_1,n_2,\cdots,n_k,n_1)$
on the graph, we can always find an eta-pairing operator $\eta^\dag$ such that the
state $(\eta^\dag)^{N}\ket{0}$ is an eigenstate of $H_m$ and $H$.

\section{States different from $(\eta^\dag)^{N}\ket{0}$}
In this section, we show that when the condition in Eq. (\ref{new}) is satisfied  for any loop $(n_1,n_2,\cdots,n_k,n_1)$
on the graph, we can find new exact eigenstates of the Hubbard-type Hamiltonian $H_m$ in Eq. (\ref{hm}) that are different from the eta-pairing state $(\eta^\dag)^{N}\ket{0}$.

When the condition in Eq. (\ref{new}) is satisfied  for any loop, as discussed in the above section we can find an $\eta^\dag$ so that there is $[H_m,\eta^\dag]=U\eta^\dag$. It will lead to the fact that
$(\eta^\dag)^{N}\ket{0}$ is an eigenstate of $H_m$.
Define $T=T_a+T_b$ with two operators
\begin{align}
&T_a=\sum_{\langle m,n\rangle}((t_{mn}+\chi_{mn})a_m^\dag a_n +H.c.), \\
&T_b=\sum_{\langle m,n\rangle}( (t_{mn}+\chi_{mn})b_m^\dag b_n +H.c.),
\overline{}
\end{align}
where $T_a$ describes the hopping of spin-up electrons and $T_b$ describes the hopping of spin-down electrons.
We can write the Hubbard-type Hamiltonian $H_m$ in Eq. (\ref{hm}) as
\begin{align}
H_m=&T +U\sum_n a_n^\dag a_n b_n^\dag b_n. \label{hm2}
\end{align}
We will show that there is
\begin{equation}
H_m (T_b)^M\eta^{\dagger}\ket{0}=0
\end{equation}
for odd integer $M$, i.e., $(T_b)^M\eta^{\dagger}\ket{0}$ is an eigenstate of $H_m$ with zero eigenvalue when $M$ is an odd integer.
This result together with $[H_m,\eta^\dag]=U\eta^\dag$ can lead to the conclusion that
\begin{equation}
\ket{\psi_{NM}}=(\eta^\dag)^{N-1} (T_b)^M\eta^{\dagger}\ket{0} \label{mn}
\end{equation}
is an eigenstate of $H_m$ with eigenvalue $(N-1)U$, where $M$ is an odd integer.
It is obvious that this new eigenstate is different from $(\eta^\dag)^{N}\ket{0}$.

Now we prove the result that there is $H_m (T_b)^M\eta^{\dagger}\ket{0}=0$ for odd integer $M$.
As the condition in Eq. (\ref{new}) is satisfied  for any loop, there is $[T,\eta^\dag]=0$, which can lead to
 $T \eta^{\dagger}\ket{0}=0$. This result together with the fact $TT_b=T_bT$
can lead to
\begin{equation}
T (T_b)^M\eta^{\dagger}\ket{0}=(T_b)^M T \eta^{\dagger}\ket{0}=0 \label{mmm}
\end{equation}
for any positive integer $M$. Notice the expression of $H_m$ in  Eq. (\ref{hm2}) and the result in Eq. (\ref{mmm}),
to show that $H_m (T_b)^M\eta^{\dagger}\ket{0}=0$ for odd $M$ we only need to show that
there is no double occupation in the state $(T_b)^M\eta^{\dagger}\ket{0}$.
Recall that $\eta^\dag=\sum_ne^{i\phi_n}a_n^\dag b_n^\dag$,
there is
\begin{equation}
(T_b)^M\eta^{\dagger}\ket{0}=\sum_n e^{i\phi_n}a_n^\dag (T_b)^M  b_n^\dag \ket{0},
\end{equation}
which means that
no double occupation in the state $(T_b)^M\eta^{\dagger}\ket{0}$ is equivalent to that $(T_b)^M  b_n^\dag \ket{0}$ is orthogonal to $b_n^\dag \ket{0}$.
When $M=1$ it is obvious that $(T_b)^M  b_n^\dag \ket{0}$ is orthogonal to $b_n^\dag \ket{0}$.
For a general integer $M$ bigger than $1$,
the state $(T_b)^M  b_n^\dag \ket{0}$ can be regarded as a spin-down electron initially at site $n$ and then hopping $M$ times.
What is the amplitude for the electron going back to its initial site $n$ after  hopping $M$ times?
When $M$ is an odd integer the way back to its starting site must contain a loop that has an odd number of sites, the electron can go back to its initial site along two opposite directions of the loop, the sum of the amplitude from these two opposite directions will be zero according to Eq. (\ref{new}).
Therefore
when $M$ is an odd integer, the state $(T_b)^M  b_n^\dag \ket{0}$ is orthogonal to $b_n^\dag \ket{0}$.

\section{Conclusions}
We have demonstrated that the condition for the \(\eta\)-pairing state \((\eta^\dagger)^N |0\rangle\) to be an eigenstate of \(H\) in Eq. (\ref{H}), which describes a Hubbard model with bond-charge interaction, is identical to the condition for it to be an eigenstate of the Hubbard-type Hamiltonian \(H_m\) in Eq. (\ref{hm}) without bond-charge interaction. For the Hamiltonian \(H\) given in Eq. (\ref{H}), we have provided the conditions on its parameters under which the \(\eta\)-pairing method can be employed to construct its exact eigenstates. Specifically, if the parameters of \(H\) in Eq. (\ref{H}) satisfy the condition in Eq. (\ref{new}) for any loop, then the \(\eta\)-pairing method becomes applicable for constructing the exact eigenstates \((\eta^\dagger)^N |0\rangle\), which are quantum many-body scar states \cite{PhysRevB.102.075132}. When the condition in Eq. (\ref{new}) is satisfied for any loop, we have shown that, in addition to \((\eta^\dagger)^N |0\rangle\), there are new eigenstates \(\ket{\psi_{NM}}\) given in Eq. (\ref{mn}) for the Hubbard-type Hamiltonian \(H_m\).
These findings improve the knowledge about Hubbard models  and the \(\eta\)-pairing method.

\section*{Acknowledgments}
We would like to thank Z.-J. Yao for helpful discussions about Hubbard models.
This work was supported by the Natural Science Foundation of Fujian Province of China (Grant No. 2022J01645).

\bibliography{ref}
\bibliographystyle{apsrev4-1}

\end{document}